\documentclass[12pt,3p,times]{article}
\usepackage{url}
\usepackage{authblk}
\usepackage[utf8]{inputenc}
\usepackage{fancyvrb}
\usepackage{tabularx}
\usepackage{caption}
\usepackage{amsmath}
\usepackage{amsthm}
\usepackage{listings}
\usepackage{geometry}
\usepackage{array}
\usepackage{xcolor}
\usepackage{tikz}
\usepackage{wasysym} 
\definecolor{myred}{RGB}{220, 50, 50} 
\definecolor{myblue}{RGB}{50, 50, 220}
\usepackage{graphicx}
\usepackage{subcaption}
\usepackage{float}

\usepackage{booktabs}

\usepackage{amssymb}

\usepackage[figuresright]{rotating}
\usepackage{tcolorbox}

\usepackage[font=footnotesize,labelfont=bf]{caption}
\usepackage[font=footnotesize,labelfont=bf]{subcaption}

\usepackage[english]{babel}
\title{
	 Advanced Scientific Methodology Plays Rossini}

\author[1]{Silvia Licciardi\thanks{Corresponding author: silvia.licciardi@unipa.it, silviakant@gmail.com, orcid 0000-0003-4564-8866.
	}} 
\author[2]{Daniela Macchione} 
\author[1]{Emmanuel Caronna} 
\author[1]{Elisa Francomano} 
\affil[1]{\normalsize University of Palermo, Department of Engineering, Viale delle Scienze, 90128, Palermo, Italy} 
\affil[2]{\normalsize Conservatory Alfredo Casella,  Via Francesco Savini, 67100, L'Aquila, Italy}

\date{}  

\begin{document}
	\maketitle

\begin{abstract}
A musical score provides the essential instructions for its performance while containing indications -- at times implicit -- regarding the composer's intentions. The presence of authorial variants, and even more so complex series of revisions associated with a single text, presents a challenging path for analytical study.
This research, situated within the application of Scientific Methodologies to Music Philology, proposes a methodological approach oriented toward the structural analysis of one of the many settings composed by Gioachino Rossini on the same Metastasio arietta ``Mi lagner\`o tacendo''.
Through Computational Analysis -- incorporating parsing, data mining, and graph theory -- the melodic, harmonic, and textual compositional choices have been rigorously explored. The results constitute a significant unicum in the field, laying the foundation for a systematic study that supports philological research and paves the way for the use of generative models to investigate the creative process.
\end{abstract}

\bigskip


\bigskip

\section{Introduction}\label{Premise}

While the scientific method finds transversal applications across many disciplines, the musical field (despite a wealth of existing studies \cite{Tymoczko2011,Cambo2006,Selway,Esparza}) has not traditionally been the primary focus of its development. In recent years, however, a growing interest in the intersections between Scientific Calculus (SC) and the arts (specifically between SC and music) has catalyzed the development of numerous computational approaches for the automated analysis of musical language (see \cite{Cambo2022,Meredith2016,Herremans2017,Seufitelli} and the refs therein). Various methodologies, including Machine Learning, Multivariate Numerical Analysis \cite{Rencher2012}, and modeling via Convolutional or Recurrent Neural Networks (CNN, RNN), have been implemented. Additionally, computational linguistics tools such as autoencoders and transformers have been employed for extracting melodic, rhythmic, or harmonic patterns, as well as for stylistic classification, authorship attribution, formal segmentation, tonal analysis, and even marketing strategies \cite{Cambo2006,Beethoven,Bach,Wei}. Most existing research, however, has focused on large-scale datasets or comparative analyses between different musical styles, adopting a macroscopic perspective of musical structure \cite{Xu2024,Zhang2021,Dervakos,Wu2023,Volk2012, White}. The present work, while situated within this broader context, represents a significant unicum. It proposes an innovative approach to the computational modeling of authorial variants, a complex field of study where even subtle differences between versions of the same piece reflect a composer's stylistic, interpretive, or developmental evolution. Ultimately, it is the creative process that drives a composer to conceive, elaborate, modify, or rewrite a musical work in its various forms.

Rather than analyzing musical style in a broad sense, this study focuses on a specific and circumscribed case: a corpus of over one hundred compositions by Gioachino Rossini based on the same text, Metastasio's arietta ``Mi lagner\`o tacendo''. Each variant possesses its own distinct character, duration, and musico-poetic features (see Section \ref{Intro}), yet all relate to a missing or implicit archetype.
In particular, this work presents a rigorous microscopic analysis of a single variant to demonstrate the methodology applied to the entire corpus. By integrating analytical tools from music data mining \cite{Conklin2002}, multivariate numerical analysis \cite{Beran2004}, graph theory \cite{Tymoczko2011,Graph}, and machine learning, alongside symbolic score analysis (e.g., via the MusicXML format \cite{Good}), a methodological synthesis capable of identifying both recurring and unique patterns within the composer's revisions is proposed. Unlike approaches that isolate individual elements such as melody, rhythm, or harmony, this study integrates multi-layered data. Specific attention is given to the text-music relationship to formally describe underlying compositional strategies. 
The results lay the foundation for a new perspective on the analysis of authorial variants, merging computational tools with classification, clustering, and predictive modeling applied to the entire set of works addressed as a cohesive whole.

\section{Computational Proposal for Gioachino Rossini's Variants}\label{Intro}
The creative process (the path that transforms an initial idea into a finished work of art) is a field of inquiry that fascinates scholars across diverse disciplines, from the humanities and sciences to neuroscience. To peer into this ``inner laboratory'', tracing the mental, technical, and material trajectories of an artist, is to touch the very core of the aesthetic experience. In this context, authorial variants (the modifications, alternatives, and revisions a creator applies to their work) represent a tangible record of thought in action, and their study serves as a privileged methodology for observing this process.
At the heart of the present research lies a case study that, by its very nature, challenges traditional philological disciplines: the vast and complex system of variants in Gioachino Rossini's (1792--1868) settings of the arietta ``Mi lagner\`o tacendo'' (MLT), from the libretto ``Siroe, re di Persia'' by Metastasio\footnote{The arietta from ``Siroe'' (Mi lagner\`o tacendo / Del mio destino avaro / Ma ch'io non t'ami, o caro, /
non lo sperar da me./ Crudele! In che t'offendo / Se resta a questo petto / Il misero diletto / Di sospirar
per te? [I will lament in silence My cruel and miserly fate; But that I do not love you, my dear, Never
hope for that from me. Cruel one! How do I offend you If in this breast remains The wretched pleasure
Of sighing for you?]) achieved considerable success and was set to music by numerous composers
besides Rossini, including, even before him, Mozart (in the Notturno K 437), Rossini's first wife,
Isabella Colbran, and her singing teacher, Girolamo Crescentini, as well as Felice Blangini. During
the same years Rossini was using it, it was also set by his friend and collaborator, Giovanni Tadolini.} (1698--1782)~\cite{Metastasio,Fabbri,Fabbri2,Rossini,Macchione,Macchione2}. With approximately one hundred autograph versions known today, composed over an extended period between the 1830s and 1860s, this corpus does not constitute a sequence of preparatory drafts for a single final work. Instead, it forms a constellation of autonomous, finished realizations. Given the notorious scarcity of sketches and preparatory materials for Rossini's grand operas, this corpus represents an invaluable document (an almost unique window into his creative workshop).
	
Rossini's settings of MLT are emblematic of a specific practice: these were not works for the general public, but musical \textit{cadeaux} (personal gifts for friends, admirers, and patrons). They were deeply rooted in 19th-century salon culture, operatic fanaticism, and the burgeoning market for autograph collecting. As early as 1842, the Parisian musical journal ``L'ind\'ependant'' noted more than one hundred such settings, a testament to the composer's prolificacy. This insistence on a single text may also reveal a personal dimension, reflecting an ``inner unease'' during a period marked by professional transition and chronic health issues. The nature of this reticulated corpus eludes traditional philological frameworks. It does not fit the model of Italian ``Authorial Philology'' as defined in \cite{Contini}, as each variant is a ``released'' work and an autonomous cadeau, rendering the notions of a ``final intention'' or unidirectional, successive choices problematic. Nor does it fully conform to French Genetic Criticism (critique g\'en\'etique), which typically analyzes the avant-texte, because in the MLT corpus, the variants are themselves the completed texts \cite{Gresillon,deBiasi}.
The MLT system acts as a distributed genetic archive, where the creative process resides not in the corrections on a single page, but in the structural relationships between multiple finished realizations. The lattice-like structure of the corpus invites an interpretation as a system analogous to a biological or artificial neural network. This metaphor is not intended to reduce Rossini's creative act (that of a conscious agent guided by aesthetics and social relations) to a mere algorithm. It offers a hermeneutic tool for decoding its internal logic.
In fact, the corpus functions as a network where a single generative idea has followed various multiplicative and transfiguring paths over decades. This structural peculiarity suggests supplementing philological hermeneutics with the analytical rigor of Scientific Calculus \cite{Quarteroni}. The central thesis of this work is that a profound understanding of such a complex phenomenon emerges from a transdisciplinary approach, where the convergence of philology, musicology, and mathematics enriches our vision of artistic creation.

\section{Hints on State of the Art}

Rossini's creativity is embodied in the physical act of writing and singing (breath, vocal production); composing is not an abstraction but a physical experience that informs creative choices. It is also enactive, as the artist does not merely transcribe a pre-formed idea but ``thinks through'' the material in a dynamic dialogue.

The state of the art is founded on the following two pillars.

\begin{enumerate}
	\item [i)] Statistical Analysis of the Musical Product: tools such as ``n-gram'' models and Information Theory quantify the syntactic regularities of music. Large-scale studies, such as the analysis by Moss et al. \cite{Beethoven} on Beethoven's quartets, have confirmed that chord frequency distributions follow a power law and that harmonic transitions are predictable. Similarly, Kulkarni et al. \cite{Bach} have used Network Science to demonstrate how Bach's works balance high informational content (entropy) with low perceptual complexity. 
	
	\item [ii)] Modeling the Creative Process via AI: AI offers various tools for investigating creativity. Using Boden's framework \cite{Boden}, which distinguishes between combinatorial, exploratory, and transformational creativity, researchers have developed models, such as Variational Autoencoders (VAEs) and Generative Adversarial Networks (GANs), capable of exploring ``latent spaces'' to generate novel yet coherent artifacts. 
\end{enumerate}

 However, these approaches primarily describe the properties of the finished product: they cannot fully illuminate the deliberative, non-linear, and contextual process that generated it. As Brandt \cite{Brandt} highlights in the comparison between Beethoven's Ninth Symphony and an AI-completed Tenth, current AI models struggle to replicate a composer's revisionist thinking. This research addresses this gap by focusing on three innovative fronts: 1) the object: authorial variants as tangible traces of the creative process rather than a single finished work; 2) the genre: vocal chamber music, where the text acts as a primary structural determinant;  3) the aim: a holistic analysis integrating melodic, harmonic, rhythmic, and interpretive-textual dimensions through Scientific Calculus.

Before moving on to the methodological practice used, in the following subsection
 let's examine two of the above-mentioned texts specifically.

\subsection{Existing Computational Modeling for Musical Composition}
In \cite{Beethoven} the authors propose a quantitative characterization of tonal
harmony through the large-scale analysis of the Annotated Beethoven Corpus (ABC)
\cite{ABC}, comprising approximately $28,000$ chord annotations extracted from digitized
string quartets. Moving beyond traditional qualitative descriptions, the study adopts
statistical modeling and distance reading techniques \cite{Moretti} to investigate harmonic
organization at scale. Chords are encoded according to degree, mode, inversion and
accidentals, providing a standardized representation suitable for computational analysis.
Tonal harmony is operationalized through four structural properties: centricity
(presence of a stable tonal center), referentiality (syntactic dependence among chords),
directionality (functional attraction toward the tonic), and hierarchy (multi-level organization
across harmonic and tonal domains). Frequency analysis reveals that a limited number of
harmonic functions dominate usage, following a Zipf-Mandelbrot distribution \cite{Tunni}
\begin{equation*}
	f(r_c)=\frac{a}{(b+r_c)^d},
\end{equation*}
where $r_c$ denotes the rank of the occurrences number of
the chord $c$ and $a,b,c$ are parameters of normalization, shift and rate of decrease of
the distribution respectively\footnote{The Zipf--Mandelbrot distribution is frequently used in
linguistic analysis because it is suitable for modeling small phenomena that occur more
frequently than large phenomena (e.g. small jumps compared to large jumps).}. Sequential
dependencies are modeled through $n$-gram statistics \cite{jurafsky2009ngrams} under a Markov assumption \cite{markov1913onegin}, $p(c_k \mid c_1,\dots,c_{k-1}) \simeq p(c_k \mid c_{k-(n-1)},\dots,c_{k-1})$,
showing that the probability of a chord $c_k$ is not given by all previous history but by a
sequence of $n-1$ chords preceding it. Transition probabilities between chords are
represented through heatmaps and quantified via entropy measures, providing a statistical
estimate of harmonic predictability. Bootstrap testing\footnote{Bootstrap testing is a
nonparametric statistical technique used to estimate the distribution of a statistic (e.g.
mean, variance, median) through repeated sampling with replacement from observed data.
Starting from an observed sample $x$ of size $n$, $N$ random samples of size $n$ are
generated by choosing random values with reinsertion from the original sample $x$. BT is
useful because it does not require assumptions about the distribution of data and, given its
flexibility, is also applicable to complex statistics.} (BT) \cite{chihara2018resampling} is employed to assess the
influence of structural features such as inversion or suspension on transition probabilities.
Directionality is evaluated through asymmetry indices between forward and reverse
transitions $\mathrm{sym}(a\rightarrow
b)=\min\!\left(\frac{p_{ab}}{p_{ba}},\frac{p_{ba}}{p_{ab}}\right)\in[0,1]$, confirming the
intrinsically non-symmetric organization of tonal harmony. Furthermore, comparison
between chord transitions and key modulations reveals distinct hierarchical rules,
suggesting that syntactic organization is not uniform across structural levels. While the
study provides a rigorous quantitative framework for tonal analysis, it is restricted to
harmonic structure within a single corpus and does not incorporate melodic, rhythmic or
textual dimensions.
A complementary perspective is presented in \cite{Bach}, where compositions of
J.S. Bach are analyzed using Network Science, Information Theory and Statistical
Physics. Each musical structure is modeled as a directed network whose nodes represent
notes and whose edges encode transitions between successive events. Information
content is quantified through the Shannon entropy \cite{Shannon} of a random walk on the
network 
\begin{equation*}
	S = -\sum_i \pi_i \sum_j P_{ij} \log P_{ij},
\end{equation*}
where $P_{ij}$ denotes transition
probabilities and $\pi_i$ the stationary distribution. For unweighted networks, local entropy
depends on the out-degree of the node $k_i$, $S_i~=~\log(k_i^{\mathrm{out}})$, linking
informational content to structural heterogeneity.
Bach's musical networks exhibit higher entropy than randomized counterparts of
equal size, indicating a structured balance between regularity and variability. Different
compositional forms cluster according to informational content, with chorales displaying
lower entropy (higher predictability) than toccatas and preludes. To model perception, the
authors introduce a cognitively constrained representation of transition structure where the
inferred transition structure $\hat{P}$ is related to the true one $P$ by the relation 
\begin{equation*}
	\hat{P} = (1-\eta)P(I-\eta P)^{-1},
\end{equation*}
where $P$ and $\eta\in[0,1]$, so capturing imperfect internal
encoding of musical organization. The efficiency of information transmission is evaluated
through the Kullback-Leibler (KL) divergence \cite{kullback1951sufficiency} 
\begin{equation*}
	D_{KL}(P\|\hat{P}) = -\sum_i \pi_i \sum_j P_{ij}\log\frac{\hat{P}_{ij}}{P_{ij}},
\end{equation*}
which is systematically lower for Bach's networks than
for random models. Introducing transition weights further reduces entropy and improves
alignment between structural and inferred representations.
Despite its methodological sophistication, this framework relies on a reduced
representation based primarily on pitch transitions, omitting other musically relevant
dimensions such as rhythm, timbre and textual content. Moreover, both studies adopt
composer specific corpora, limiting generalizability across styles and historical contexts.

Overall, these contributions demonstrate the effectiveness of statistical modeling,
network representations and information theoretic measures in computational musicology.
However, they do not address the integrated analysis of melodic, harmonic, rhythmic and
textual components. The present work aims to fill this gap by proposing a general
computational framework for joint melodic-textual and harmonic-textual analysis,
applicable beyond the specific Rossinian case study considered here.

\section{Dataset and Methodology}\label{DM}
	
The musical sources of ``Mi lagner\`o tacendo'', identified and edited by Macchione in \cite{Macchione2},constitute the primary resource of this research. The corpus consists of 133 scores available in ``.sib'' format (Sibelius) grouped into nineteen ``types'' or main families of variants. Rossini used Metastasio's verses \textit{``Mi lagner\`o tacendo della mia sorte amara ma ch'io non t'ami, o cara, non lo sperar da me''} as a fixed theme, weaving infinite musical transfigurations that range from lyrical canzonette to energetic boleros. These settings represent a deliberate exploration of musical character, where the composer encodes distinct emotional and technical profiles (from joy, irony to melancholy and sorrow) into each version, effectively treating the single text as a multifaceted laboratory for stylistic and interpretive experimentation. To decipher this complex system, traditional musicology can turn to Ruwet's paradigmatic analysis \cite{Ruwet} or Genette's theories on transtextuality \cite{Genette}. However, musicology alone lacks the tools to automate the extraction and processing of such multi-layered data. The integration of Scientific Calculus does not replace musical hermeneutics but enhances it, providing the rigor necessary to manage otherwise intractable complexity.
This approach translates qualitative hypotheses into quantifiable problems.

\begin{figure}[h!]
	\centering
	\includegraphics[width=0.6\linewidth]{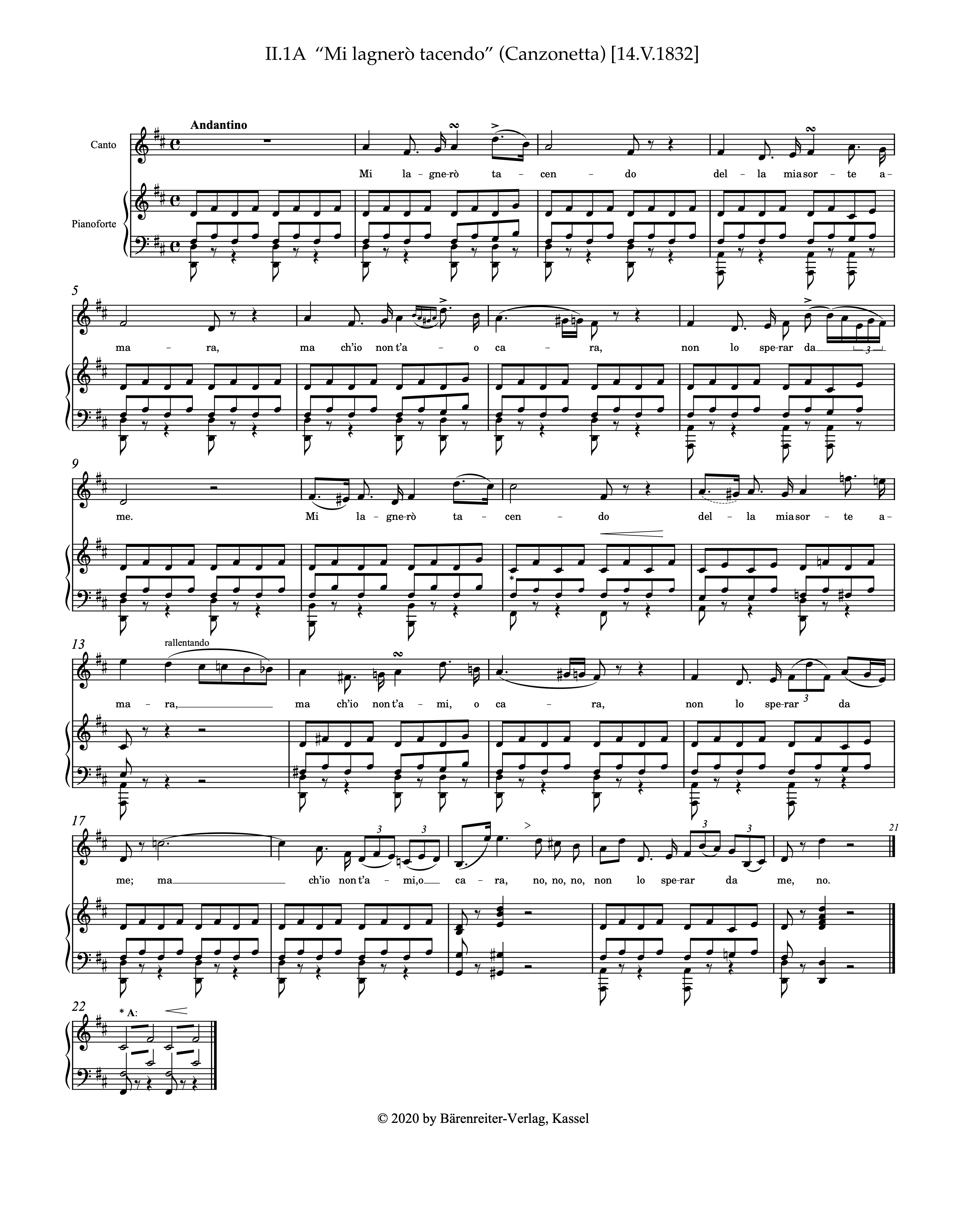}
	\caption{Score of variant II.1A in \cite{Macchione2}.}
	\label{primavar}
\end{figure}

As a test case, the variant identified as II.1A (belonging to one of the oldest families of such pieces) in \cite{Macchione2} has been selected, for voice and piano (see Fig.~\ref{primavar}). The initial process involved ``parsing'', i.e. sequentially reading the file elements to extract and organize relevant data into computationally interpretable structures.
This study utilized both Matlab \cite{MATLAB} and Python \cite{Python} in parallel to validate results and compare algorithmic performance. Ultimately, this methodology shifts the emphasis from the analysis of static musical syntax to the dynamic modeling of the creative thought process.

\subsection{Parsing}
Parsing is the fundamental process by which the raw digital representation of a musical score is analyzed and decomposed into its syntactic constituents. This step is crucial to bridge the gap between human-readable notation and machine-readable structures, enabling the extraction of structured data suitable for statistical analysis and algorithmic processing.

\begin{figure}[h!]
	\centering
	\includegraphics[width=1.0\textwidth]{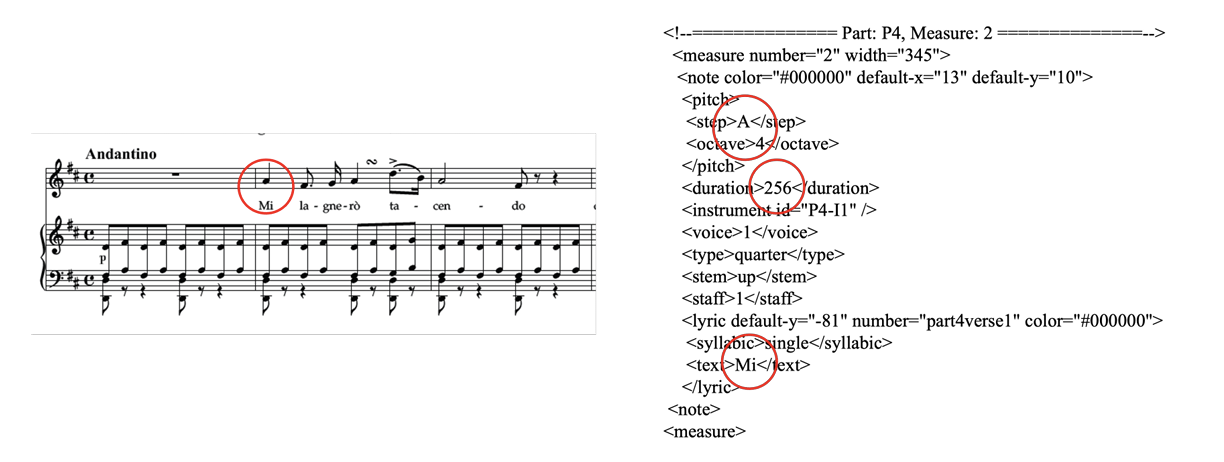} 
	\caption{From graphical notation to XML encoding. The red circles highlight the correspondence between the visual note associated with the syllable `Mi' and its specific tags: the pitch coordinates (\texttt{<step>A</step>}, \texttt{<octave>4</octave>}), the temporal value (\texttt{<duration>}), and the textual content (\texttt{<text>Mi</text>}).}
	\label{fig:spartitoxml}
\end{figure}

The primary source material for this study consists of digital scores encoded in ``MusicXML'' format \cite{Good}. MusicXML is the standard open format for exchanging digital sheet music; it represents the score as a hierarchical tree of XML tags, separating logical content (pitch, rhythm) from layout information.
To illustrate this translation from graphical symbols to code, Fig.~\ref{fig:spartitoxml} provides a comparative visualization. The figure highlights a specific musical event, the opening note of the vocal line corresponding to the lyric `\textit{Mi}', and its underlying XML descriptor.

As shown in the red circles in Fig.~\ref{fig:spartitoxml}, the parser navigates the XML tree to retrieve discrete attributes for every event (see Appendix \ref{Appendix - Music Fundamentals}).
\begin{itemize}
	\item \textbf{Pitch definition}: The visual note located on the second space of the treble clef is encoded via the \texttt{<pitch>} tag\footnote{In the musical field, a note's position on the staff is defined as `pitch', denoted by its name (e.g., A in Germanic notation or La in the fixed-do system) and its specific octave (e.g., A4 for the central octave). In the scientific domain it refers to `frequency' ($f$) to define the corresponding acoustic signal (e.g., A4 = 440 Hz). Given the transdisciplinary nature of this work, both terminologies will be employed as required by the analytical context (see Appendix \ref{Appendix - Music Fundamentals} for further technical specifications).}, which splits the information into the diatonic step (\texttt{<step>A</step>}) and the octave index (\texttt{<octave>4</octave>}).
	\item \textbf{Temporal definition}: The rhythmic value (a quarter note) is converted into a numerical integer within the \texttt{<duration>} tag (e.g., 256), which represents the note's length in terms of ``divisions'' per quarter note.
	\item \textbf{Lyric association}: The syllable `\textit{Mi}' written below the staff is encapsulated in the \texttt{<lyric>} block within the \texttt{<text>} tag. This allows the system to semantically link the acoustic event (Note A4) with the linguistic event (Syllable `Mi').
\end{itemize}

The organization of this type of information can be carried out in different ways depending on the type of analysis to be performed. In Matlab, for example, a common representation is the one based on structures (\texttt{struct}), in which each musical element, such as a note, is encoded via explicit fields\footnote{The previous lines become e.g.: \texttt{note.pitch = 'A4'; note.duration = 256; note.type = 'quarter'; note.stem = 'up'; note.voice = 1; note.staff = 1; note.lyric.text = 'Mi'; note.lyric.syllabic = 'single'.}}. This form is useful for programmatic access to single attributes, but does not constitute the only possible transcription of XML data. The MusicXML format, in fact, is a tree structure, of hierarchical type, non-sequential, and lends itself to multiple decoding modes. Reading the file via xmlread (or similar libraries like xmlstruct) yields a DOM (Document Object Model), allowing the data to be reinterpreted as tables (\texttt{table}), vectors of objects, or nodes and edges within a directed graph (\texttt{digraph}). The choice depends on the goal of the analysis: for studies of statistical or sequential type a table can be useful, while for structural or relational analyses (e.g. melodic flows or text-music associations as in our objectives) a representation through Directed Graphs \cite{DIestel} can turn out to be more effective. The peculiarity of a musical score is in fact the proceeding in time both rhythmically and melodically/harmonically, including however a contemporaneity of events (melody, chords, text, instruments), for which the creation of nodes and edges results in an implementation choice that allows organizing the information in an accessible way.

\subsubsection{Graph Topology: Node and Edge Definitions}
\begin{figure}[h!]
    \centering
    \includegraphics[width=0.8\textwidth]{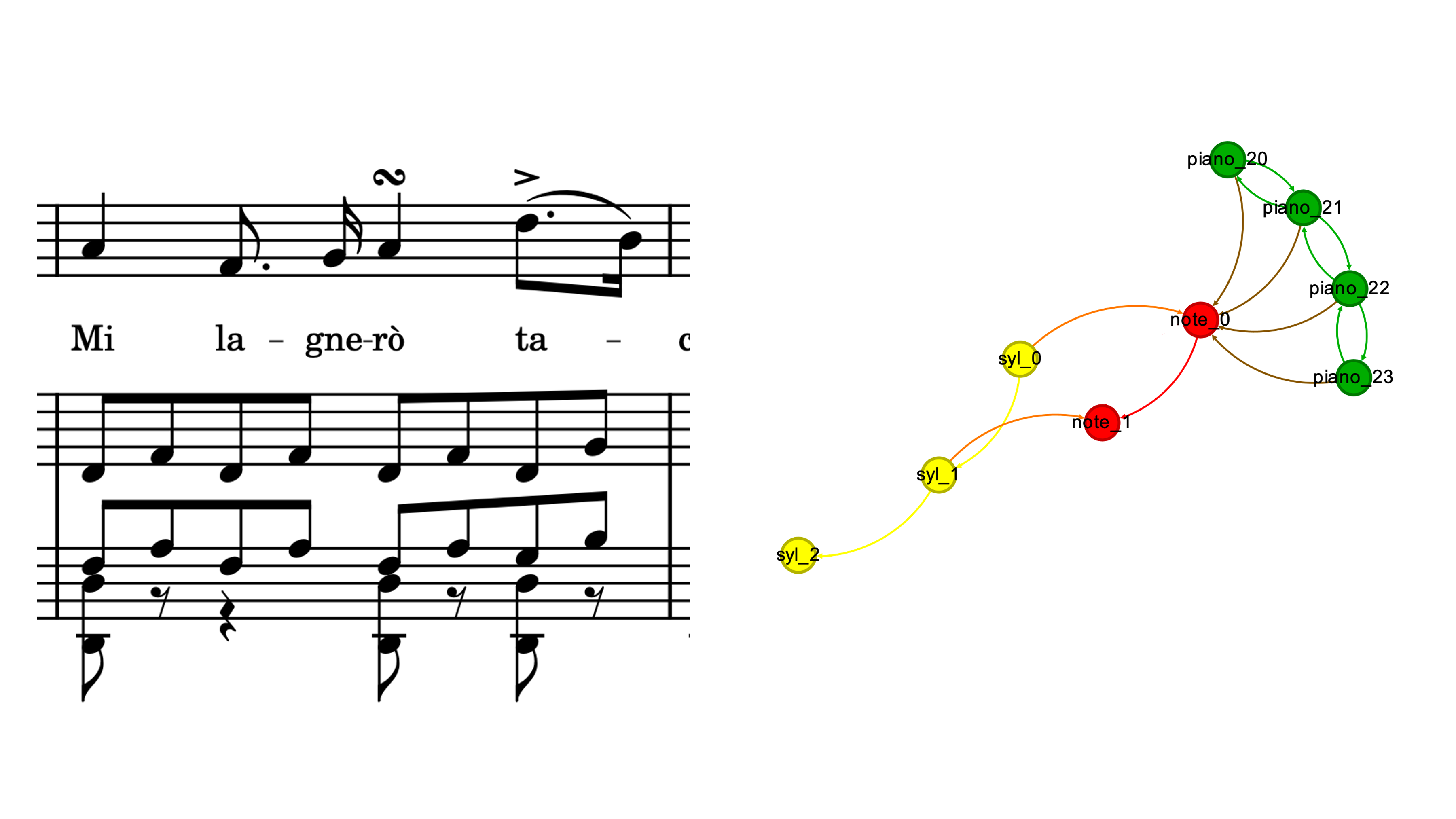} 
    \caption{Graph structure of the second measure of the variant.}
    \label{fig:graphstructure}
\end{figure}

The proposed model is realized as a heterogeneous graph (see Fig.~\ref{fig:graphstructure}) that integrates three distinct information domains: the melodic line (voice), the textual content (lyrics), and the harmonic support (piano). Below, the node types and the relational structures (edges) that constitute the graph's architecture have been formally defined.

\paragraph{Node Types}
Nodes represent the fundamental entities of the musical score. Each node type is equipped with specific feature sets designed to ensure consistency and comparability:

\begin{itemize}
    \item \textbf{\texttt{note}}: these nodes represent the sonic events of the main vocal part. Each node encodes acoustic-musical features (MIDI pitch, quantized duration), metric details (normalized position within the measure), and theoretical attributes (local tonal function). A boolean attribute distinguishes notes that serve as the ``head'' of a syllable from those that constitute melismatic prolongations.
    
    \item \textbf{\texttt{syllable}}: these nodes represent the textual units extracted from the lyrics. Their existence is conditional on the presence of text associated with the music; these nodes act as semantic anchors for prosodic analysis.
    
    \item \textbf{\texttt{piano}}: these nodes represent the sonic events of the accompaniment. To maintain homogeneity with the vocal line, these nodes inherit the exact same feature schema as the \texttt{note} nodes (continuous vectors and harmonic identifiers), facilitating direct comparative analysis between melody and harmony. Polyphonic chords are ``exploded'' into individual nodes to allow for granular management of simultaneous voices.
\end{itemize}

\paragraph{Edge Types}
The relationships between nodes define the temporal, vertical, and semantic structure of the piece. The edges are categorized into three functional families:

\begin{itemize}
    \item \textbf{Sequential Relationships (Horizontal Time)}:
    These define the chronological order of events within a single modality.
    \begin{itemize}
        \item \texttt{(node note, edge next\_note, note)}: connects melodic notes in succession, reconstructing the linear melodic flow;
        \item \texttt{(syllable, next\_syllable, syllable)}: connects syllables according to the reading order, preserving the syntactic structure of the text;
        \item \texttt{(piano, next, piano)}: connects successive onsets in the accompaniment, defining the rhythmic-harmonic skeleton of the piece.
    \end{itemize}

    \item \textbf{Structural Relationships (Vertical Time)}:
    These model the simultaneity of events.
    \begin{itemize}
        \item \texttt{(piano, vert, piano)}: Connects notes playing simultaneously (chords) within the piano part based on onset bucketing. This is an undirected relationship that allows for information exchange between co-occurring harmonic voices.
    \end{itemize}

    \item \textbf{Alignment Relationships (Cross-Modal)}:
    These connect different domains, synchronizing text, melody, and harmony.
    \begin{itemize}
        \item \texttt{(syllable, sung\_on\_head, note)}: connects a syllable exclusively to the initial note (head) of the corresponding melisma. This is the fundamental relationship for precise prosodic alignment;
        \item \texttt{(syllable, sung\_on, note)}: extends the previous relationship by connecting the syllable to \emph{all} notes comprising the melisma (representing the full temporal span of the syllable);
        \item \texttt{(piano, sung\_on, note)}: realizes the synchronization between accompaniment and vocals. This edge connects a piano note to a vocal note if and only if their onset times coincide (within a computational tolerance $\varepsilon$). This provides an informative bridge between the harmonic voicing and the melodic pitch.
    \end{itemize}
\end{itemize}

\begin{figure}[h!]
    \centering
    \includegraphics[width=0.5\textwidth]{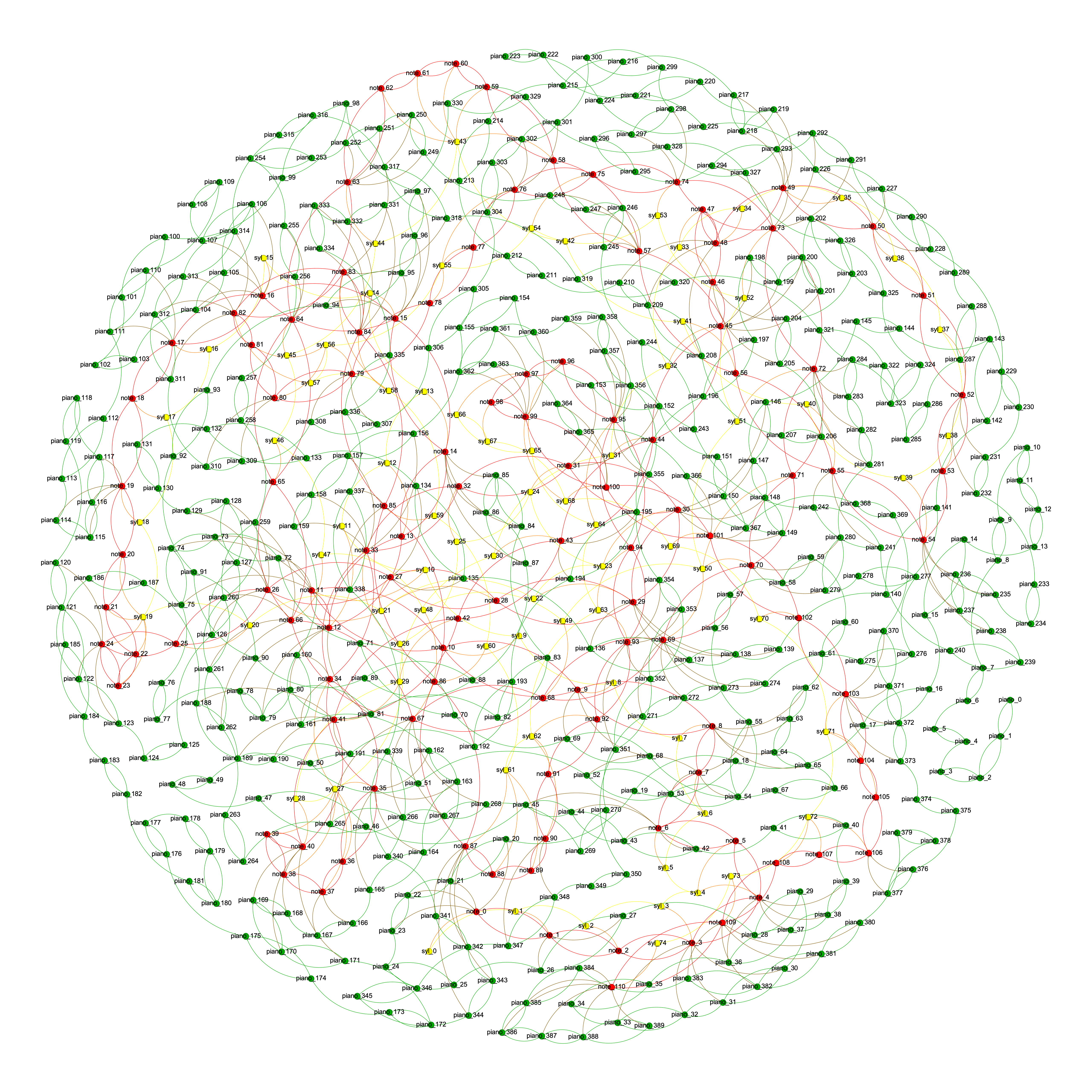} 
    \caption{Complete heterogeneous graph representation of the entire score.}
    \label{fig:full_graph}
\end{figure}

As previously introduced, representing a score as a heterogeneous graph allows to model simultaneously the \emph{melodic} dimension (notes) and the \emph{linguistic} one (syllables), capture multiple alignments (e.g. melisma or notes repeated on the same syllable), and exploit the local \emph{context} through paths of variable length. This approach is useful to classify the typology of notes, predict phrasings, or generate melodies coherent with the style. The resulting complex network, representing the entire score, is visualized in Fig.~\ref{fig:full_graph}.

\section{Data Analysis}
Once the extraction of data via parsing is completed and the graph modeling is performed, the data are analyzed in accordance with the following characteristics, which are useful for several analyses of the score.

\begin{enumerate}
	\item \textbf{Melodic and Textual Analysis}: 
	\begin{enumerate}
		\item syllabic distribution (mapping the distribution of individual syllables across the melodic line to identify the passage's character (e.g., melismatic/virtuosic vs. syllabic/lyrical); 
		\item intervallic analysis (identification of specific lexical units emphasized by melodic leaps or chromaticism);
		\item rhythmic-textual correlation (analysis of note durations in relation to syllabification to determine how textual prosody shapes rhythmic structures); 
		\item dynamic profiling (investigation of dynamic markings as a function of textual emphasis or melodic contour)
		\item ornamental syntax (analysis of the use and placement of ornamentation in relation to the poetic structure).
	\end{enumerate}
	
	\item \textbf{Harmonic and Textual Analysis}: 
	\begin{enumerate}
		\item harmonic rhythm (evaluation of the rate of harmonic change and its link to textual meaning (e.g., the use of abrupt modulations to highlight structural or emotional shifts in the text)); 
		\item vertical color and dissonance (analysis of chromatic chords and dissonances in correspondence with ``keywords'' to reinforce high-intensity emotional states); 
		\item consonance and tension modeling (quantification of the degree of consonance and harmonic tension to define the specific expressive character of each musical phrase).
	\end{enumerate}
\end{enumerate}

\noindent The combined analysis of these aspects allows us to highlight recurrent compositional strategies or significant differences between the variants of the MLT corpus, offering a deeper interpretation of the musical evolution of the piece and/or of the pieces in their entirety.

\subsection{Melodic-Textual Analysis of the Extracted Data}\label{MTA}

After a first extraction of the characteristics of the melodic line and of the textual line, in Table~\ref{TabDatiCombo} the information derived from the combined melody -- text graph is summarized, according to the indicators reported below:
\begin{table}[h!]
\centering
	\caption{Extraction Data on the Melody -- Lyrics Graph}\label{TabDatiCombo}
	\begin{tabular}{@{} l r @{}}
		\toprule
		Melody-Lyrics Graph Data & {\#} \\
		\midrule
		Melody Nodes & 111 \\
		Lyric Nodes & 75 \\
		Total Nodes & 186 \\
		Edges notes $\to$ syllables & 111 \\
		Total Edges & 295 \\
		Graph Density $\delta_G$ & 0.0086\\
		Average In-degree (notes) & 0.99 \\
		Average Out-degree (notes) & 1.99 \\
		Average In-degree (syllables) & 2.47 \\
		Average Out-degree (syllables) & 0.99 \\
		Max Degree & 7 \\
		Min Degree & 0 \\
		\bottomrule
	\end{tabular}
\end{table}

\begin{figure}[h!]
	\centering
	\includegraphics[width=1\linewidth]{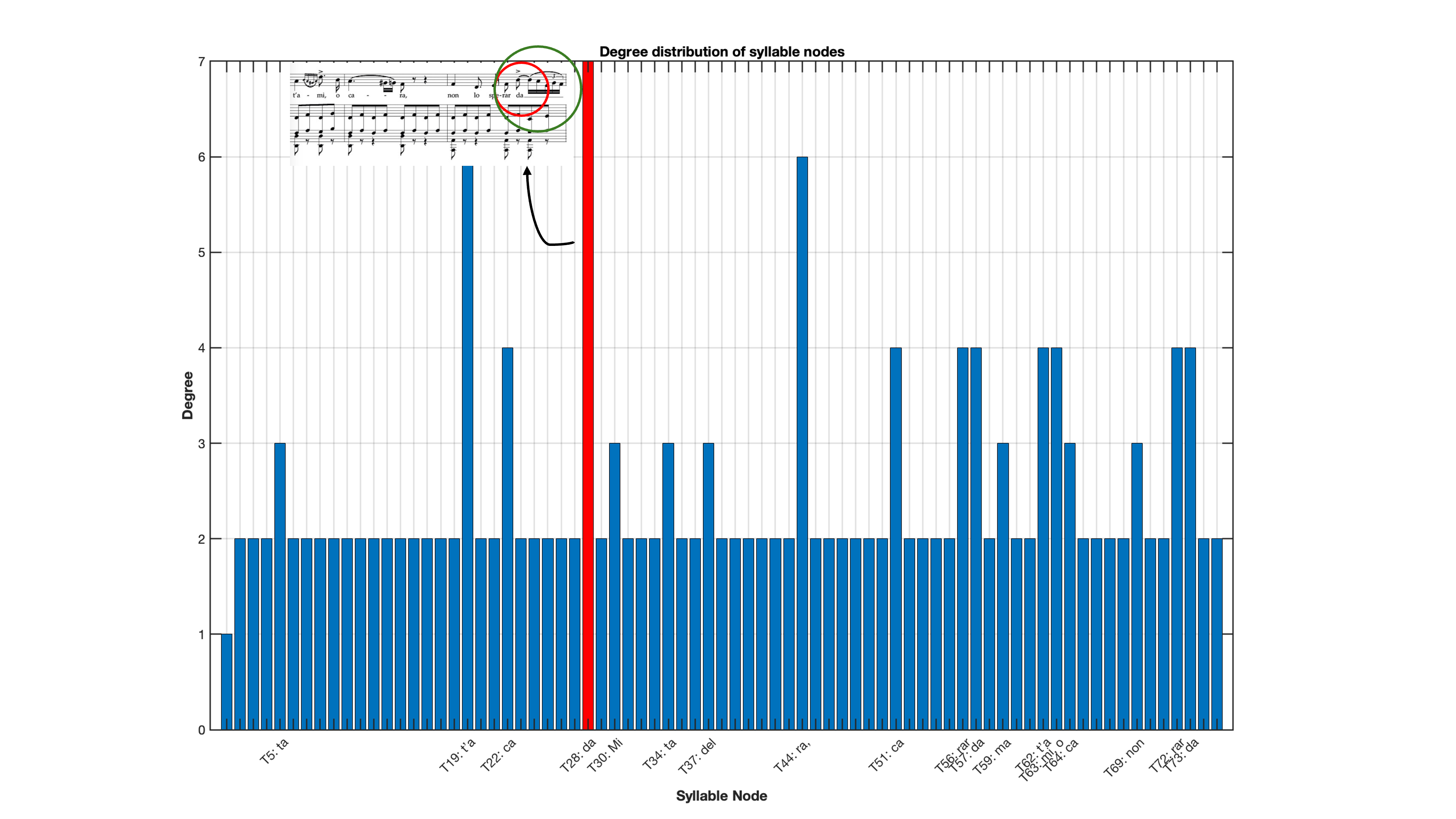}
	\caption{Histogram of the degree in of the syllable nodes with the max degree in red. The labels reported regard only the degrees $> 2$.}\label{maxgrade}
\end{figure}

\begin{enumerate}
	\item average in and out degree represent the average of the incoming or outgoing edges in every single node (notes and syllables);
	
	\item average in-degree of the notes 0.99 is consistent with the fact that each note receives on average 1 edge (except the first); each note is in fact connected only to the immediately preceding and immediately succeeding note (melodic succession);
	
	\item average out-degree of the notes 1.99 indicates that each note emits on average about 2 edges. This is plausible because each note emits an edge towards the next note (melody) and an edge towards its associated syllable;
	\item average in-degree of the syllables 2.47 reflects the musical structure in which some notes can share the same syllable (melisma) (in Fig.~\ref{maxgrade} is shown the density of the syllable-nodes based on the number of note-nodes to which they are connected); each syllable furthermore, like the notes, is also connected to the preceding syllable and to the succeeding one;
	\item average out-degree of the syllables 0.99, as one expects, denotes that each
	syllable emits on average 1 edge (except the last);
	
	\item the density of a directed graph is a measure that expresses how much a graph is `complete` in terms of edges, with respect to the maximum possible number of edges, that is
	\begin{equation*}\label{key}
	\delta_G = \dfrac{\alpha_{tot}}{N_{tot}(N_{tot}-1)}
	\end{equation*}
	with $\alpha_{tot}$ the total number of edges present in the graph and $N_{tot}$ total number of nodes (or vertices) in the graph;

	\item max degree indicates the maximum number of connections of a specific node. In particular \textit{max degree} = 7 corresponds to the node syllable `da` connected to the following syllable node and to the 6 nodes note of the melisma. In Fig.~\ref{maxgrade} the histogram describes the degree in of the syllable nodes and it is possible to observe the syllable node with the Max Degree according to the correspondence with the shown score (only the labels with degree $> 2$ are reported because they are more informative).
\end{enumerate}

The first findings on the melodic textual analysis have provided the following results:

\begin{enumerate}
	\item \textbf{syllabic patterns}: the analysis of the syllabic patterns, connected to the different frequencies of the corresponding notes allows the identification of recurrent melodic patterns. In this regard in Fig.~\ref{PatternSillabiciPitch} is reported the distribution of the 15 most recurrent syllabic bigrams connected to melodic patterns classified according to the following typologies (see Appendix \ref{Appendix - Music Fundamentals}):
	
	\begin{itemize}
		\item \texttt{same}: same initial and final pitch;
		\item \texttt{up\_step} / \texttt{down\_step}: intervals of a second;
		\item \texttt{up\_leap} / \texttt{down\_leap}: intervals greater than a second.
	\end{itemize}

	\begin{figure}[h!]
		\centering
		\includegraphics[width=0.8\linewidth]{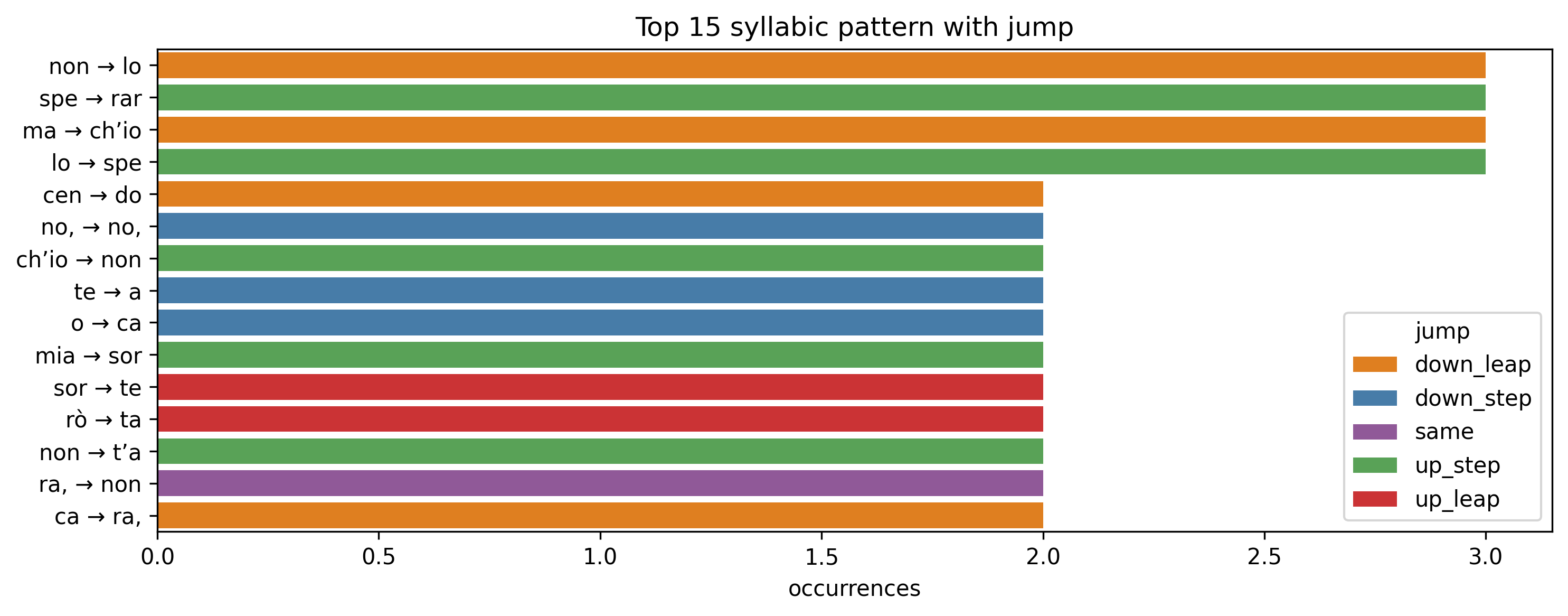}
		\caption{Frequency of the 15 most recurrent syllabic bigrams combined with pitch jump information.}
		\label{PatternSillabiciPitch}
	\end{figure}
	
	\item \textbf{frequency - duration}: a second analysis concerns the correspondence between the duration and the frequency of every single note. For simplicity the frequencies have been divided into three pitch bands (low, mid, high) and three duration bands (short, medium, long) in accordance with the Table \ref{table pitch duration}. A promising avenue for future research involves applying Multiresolution Analysis (MRA) to refine this stage of the process \cite{MRA}.\\
    
	\begin{table}[h!]
    \centering
		\caption{Pitch and duration bands definitions.}\label{table pitch duration}
		\begin{tabular}{@{} l r @{}}
			\toprule
			Pitch (0-127) & Duration (4/4) \\
			\midrule
			\textit{low} : MIDI $<$ 60 (octave below middle Do (C4)) & \textit{short} : $\le 0.5$ quarters\\
			\textit{mid} : 60 $\le$ MIDI $<$ 72 (octave between middle Do (C4) and high Do (C5)) & \textit{medium} : $\le 1.0$ quarters\\
			\textit{high} : MIDI $\ge$ 72 (octave over high Do (C5)) & \textit{long} : $> 1.0$ quarters\\
			\bottomrule
		\end{tabular}
	\end{table}

	Fig.~\ref{PitchDurata} shows this distribution in which is highlighted a greater incidence in the pairs mid-short and mid-medium, in fact:
	\begin{figure}[h!]
		\centering
		\includegraphics[width=0.6\linewidth]{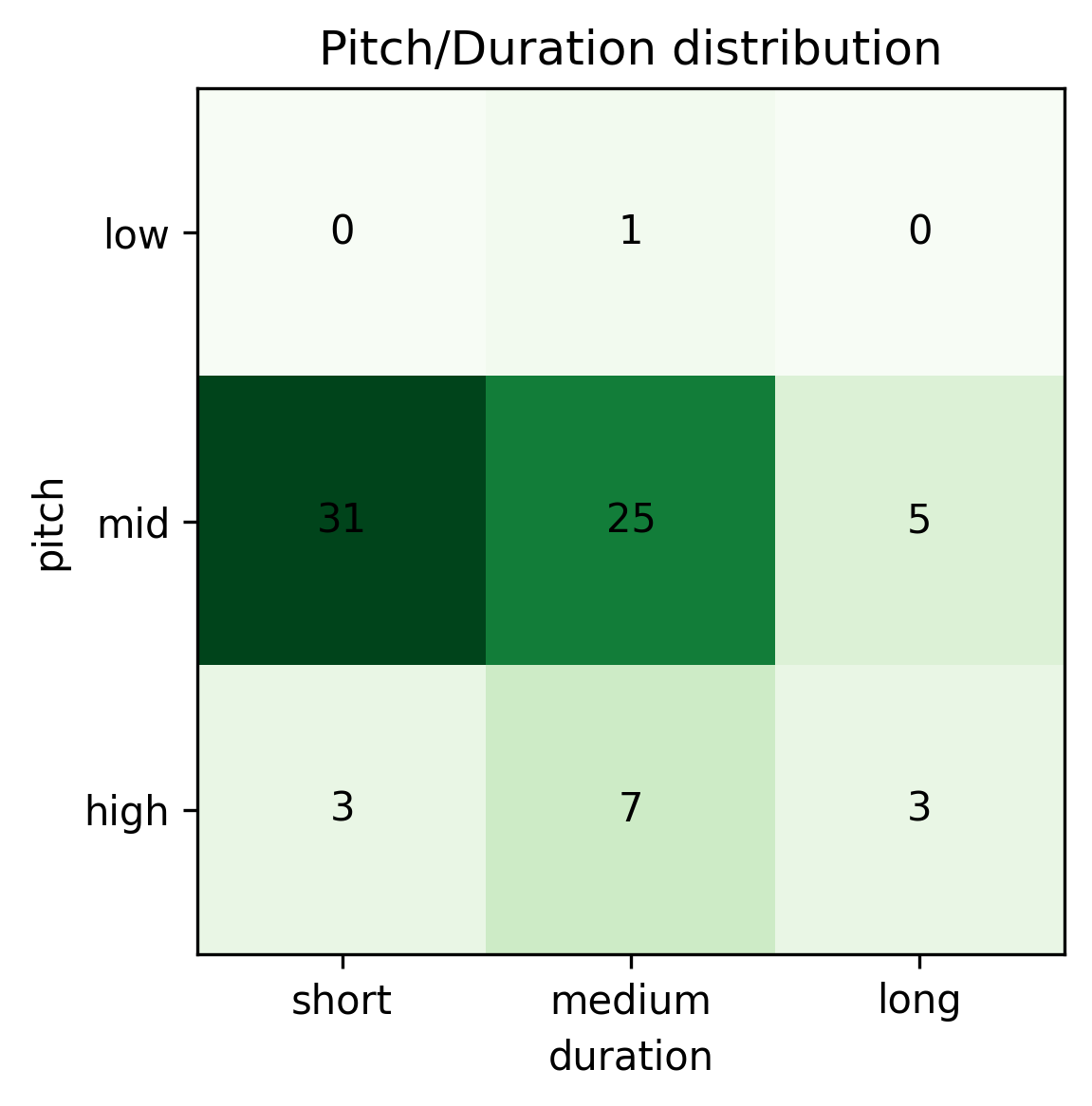}
		\caption{Pitch distribution - duration. Numerical values indicate the number of syllabic events falling into each combination. The intensity of the colour increases proportionally to the number.}
		\label{PitchDurata}
	\end{figure}
	The cell \texttt{mid/short} dominates with 30 occurrences: the majority of the syllables is sung on notes of medium register with values equal to or lower than the eighth (note), the combination \texttt{mid/medium} (25) is the second most frequent: the base pulsation falls again
	in the medium register, the \texttt{high} notes appear 13 times in total, more often with \texttt{medium} (7) than with \texttt{short} (3): the high notes enjoy slightly more extended durations and the \texttt{long} durations are rare (8 cases): they are concentrated on the
	medium and high register, consistently with points of climax or phrase closures.

	\item \textbf{Vocal distribution by Pitch-Duration}: Fig.~\ref{DurateVocali} shows the distribution of the temporal duration of every main vowel (considering also the diphthongs) of the corresponding syllable. The box represents the range between the 2$^{\text{nd}}$ and 3$^{\text{rd}}$ quartile (IQR); the line in the center, the median; the so-called whiskers, the values within $1.5 \cdot IQR$; the points outside are outliers, that is anomalous durations. The results provide the following statistics:
	\begin{description}
		\item[\texttt{a}] - median 0.75 quarters (dotted eighth). Two distinct upper outliers are noticeable: one at 2.0 and an extreme one at 3.0, suggesting notes held for a long time on the open vowel;
		\item[\texttt{e}] - median 0.62 quarters; distribution similar to \texttt{a} but slightly shorter overall; a single outlier is present at 2.0 quarters, indicating an isolated lengthening;
		\item[\texttt{i}] - median 0.75 quarters; although the median is identical to \texttt{a}, the distribution lacks extreme outliers, remaining within a range between approximately 0.1 and 1.0;
		\item[\texttt{o}] - median 0.50 quarters (eighth note); the box extends upwards from the median value (up to 1.0), indicating a tendency towards shorter durations compared to the other main vowels, often associated with phrase endings or appoggiaturas;
		\item[\texttt{u}] - absent in the text, no measure available;
		\item[\texttt{io}] - median fixed at 0.75 quarters; the absence of a box indicates zero variance (all occurrences have identical duration);
		\item[\texttt{ia}] - median fixed at 0.25 quarters; zero variance here as well, indicating a rapid and constant execution (sixteenth note) for this diphthong.
	\end{description}	
	\begin{figure}[h!]
		\centering
		\includegraphics[width=0.8\linewidth]{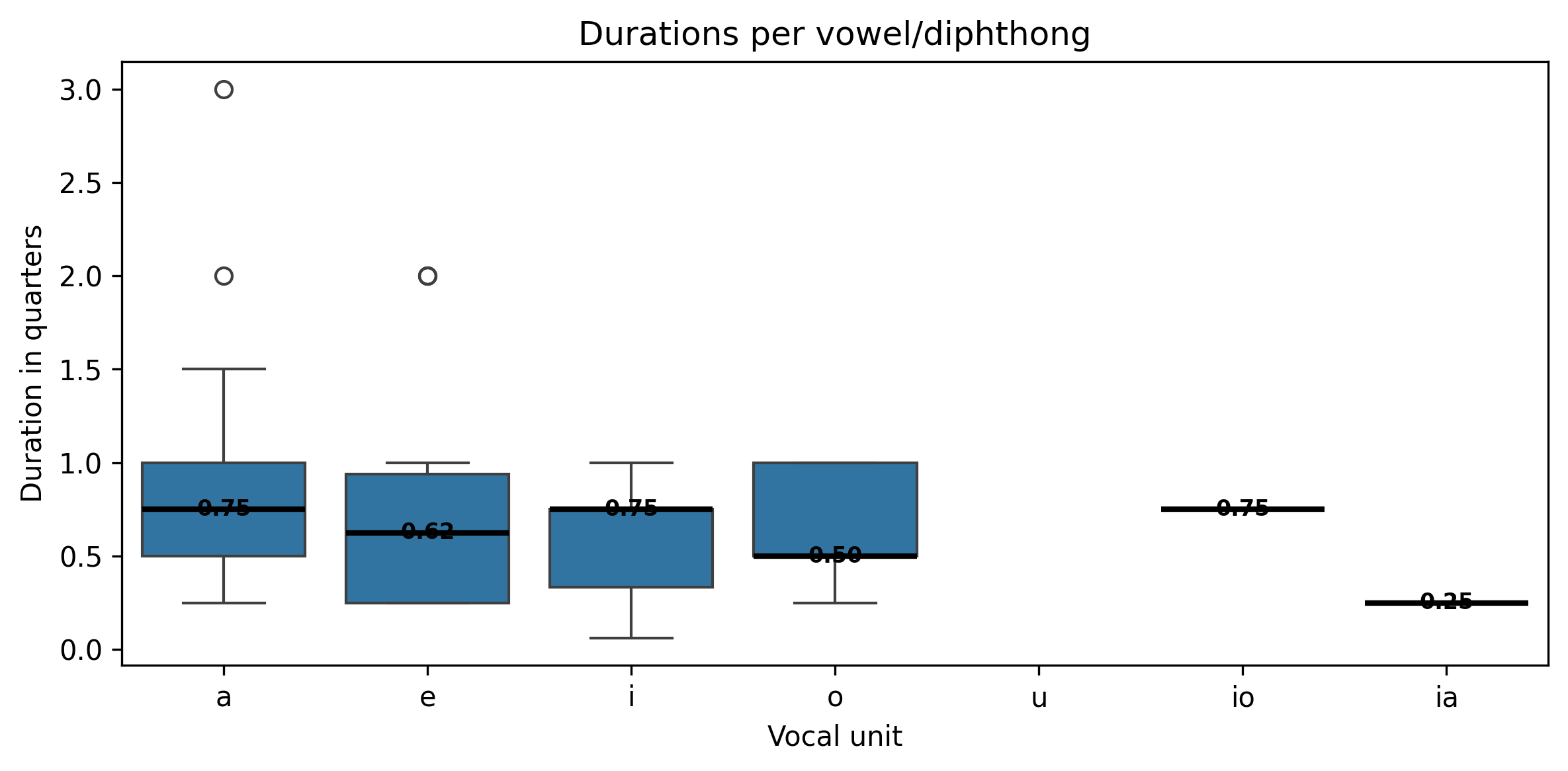}
		\caption{Duration (in quarters) distribution, grouped by principal vowel of the syllable. Circles indicate outliers. Labels indicate the median.}
		\label{DurateVocali}
	\end{figure}
	
	\item \textbf{Vowel-Vowel Transitions}:
	the distribution of the passages between contiguous vowels\footnote{All combinations were visualized to have access also to rare events but it is possible to select a minimum threshold, e.g. a filter s.t. for $x=occurrence$ let $x\ge 5$, thus eliminating the rare combinations
		(\emph{u}, diphthongs, only consonant),
		maintaining the matrix readable and concentrated on the recurrent trends in the vocal line. In this way the graph would provide, at a glance, a ``flow map'' of the vowels inside the melody, useful both for
		prosodic analysis and for possible models of synthesis or prediction of text--music. 
	} (included in the relative syllables) is shown in Fig.~\ref{TransizVoc}. The value in each cell represents the absolute count; the intensity of the red increases with the number of occurrences. The sequences $a,e\rightarrow a$, $a\rightarrow e$ and $o\rightarrow a,e$ have greater frequency; the vowel `a' is the most repeated and consequently the most used to pass to another syllable while the `i' appears less frequently and it is followed above all by the `o'. The phonetic flow privileges the open vowels.
	\begin{figure}[h!]
		\centering
		\includegraphics[width=0.5\linewidth]{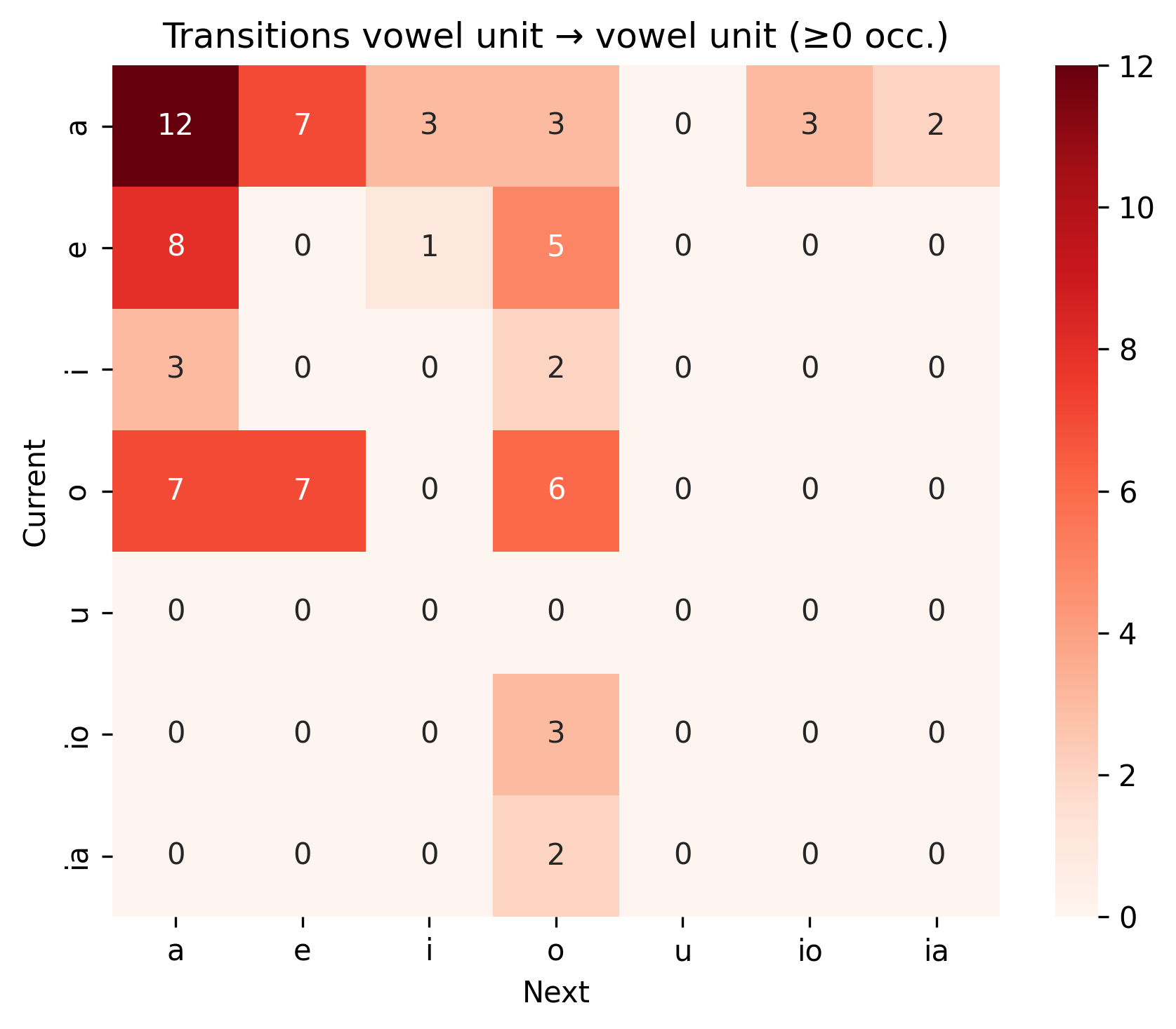}
		\caption{Distribution of the succession of vowels contained in adjacent syllables. Current syllable vowel vs next syllable vowel.}
		\label{TransizVoc}
	\end{figure}
	
\end{enumerate}	

The results shown are only a selection among the multiple information, statistics or distributions that can be highlighted from the data. Once the dataset is decoded, in fact, the extraction is automatic and depends only on the necessary requests\footnote{The graphs concerning the harmonic-textual distributions reproduce the same trends previously visualized and are not reported for brevity.}.

\subsection{Cross-Variant Statistical Analysis}\label{CVSA}

To fully exploit the potential of the computational framework and validate the hypotheses formulated on the single score, the methodology was extended to the entire corpus of the \textit{Mi lagner\`o tacendo} variants. This scaling allows for a shift from a micro-analytical perspective to a macro-analytical one, highlighting Rossini's invariant compositional strategies alongside his extreme stylistic variability. 

As a demonstrative example of this approach, an analysis conducted on a randomly selected subset of three variants is presented. Comparing the individual distributions with the aggregated data reveals how the composer's localized choices contribute to a global statistical behavior. For illustrative purposes, only the case concerning the frequency-duration distribution is presented.

\begin{figure}[h!]
        \centering
        \begin{subfigure}[b]{0.32\textwidth}
            \centering
            \includegraphics[width=\textwidth]{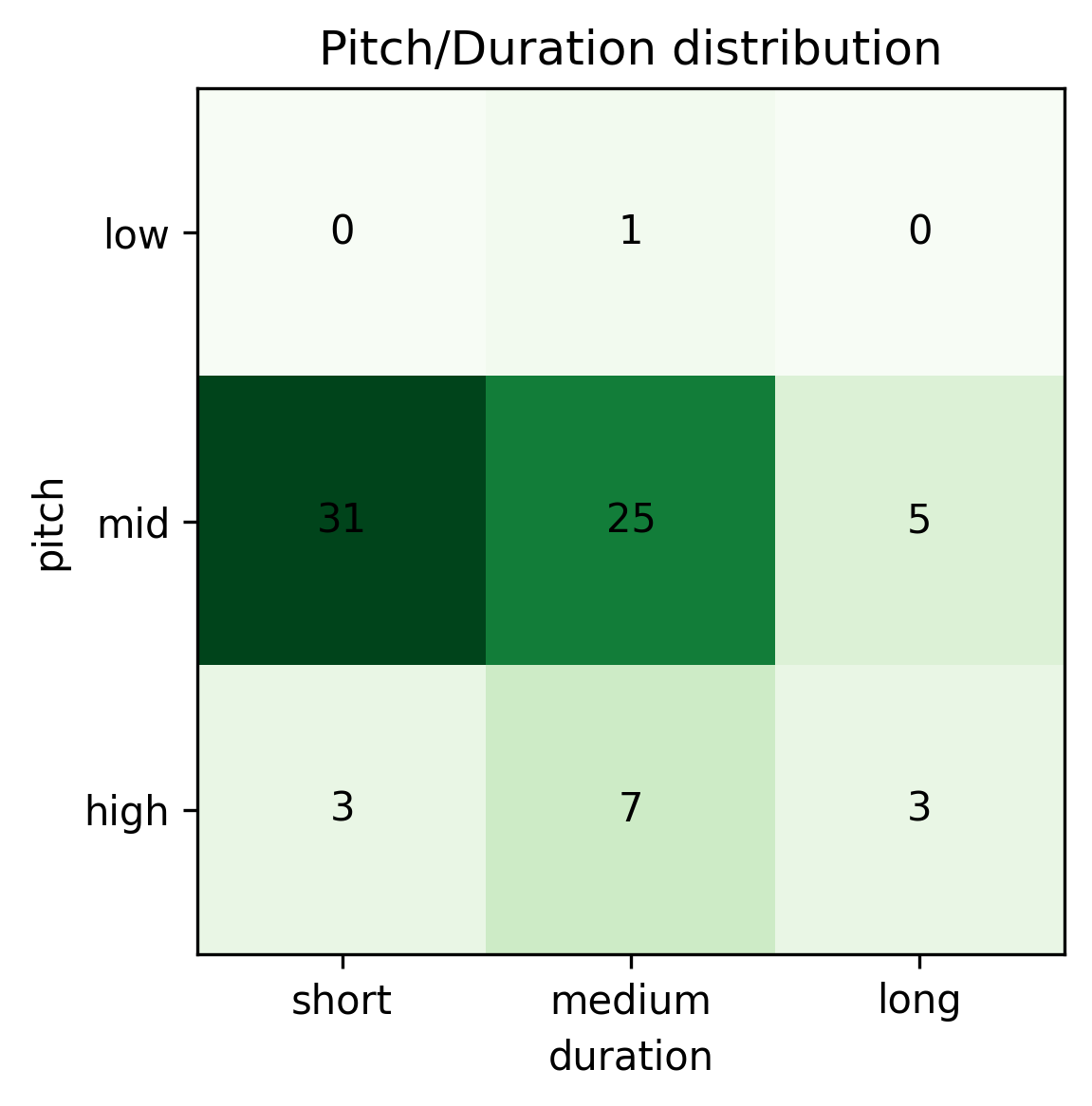}
            \caption{Variant A}
            \label{fig:var1}
        \end{subfigure}
        \hfill
        \begin{subfigure}[b]{0.32\textwidth}
            \centering
            \includegraphics[width=\textwidth]{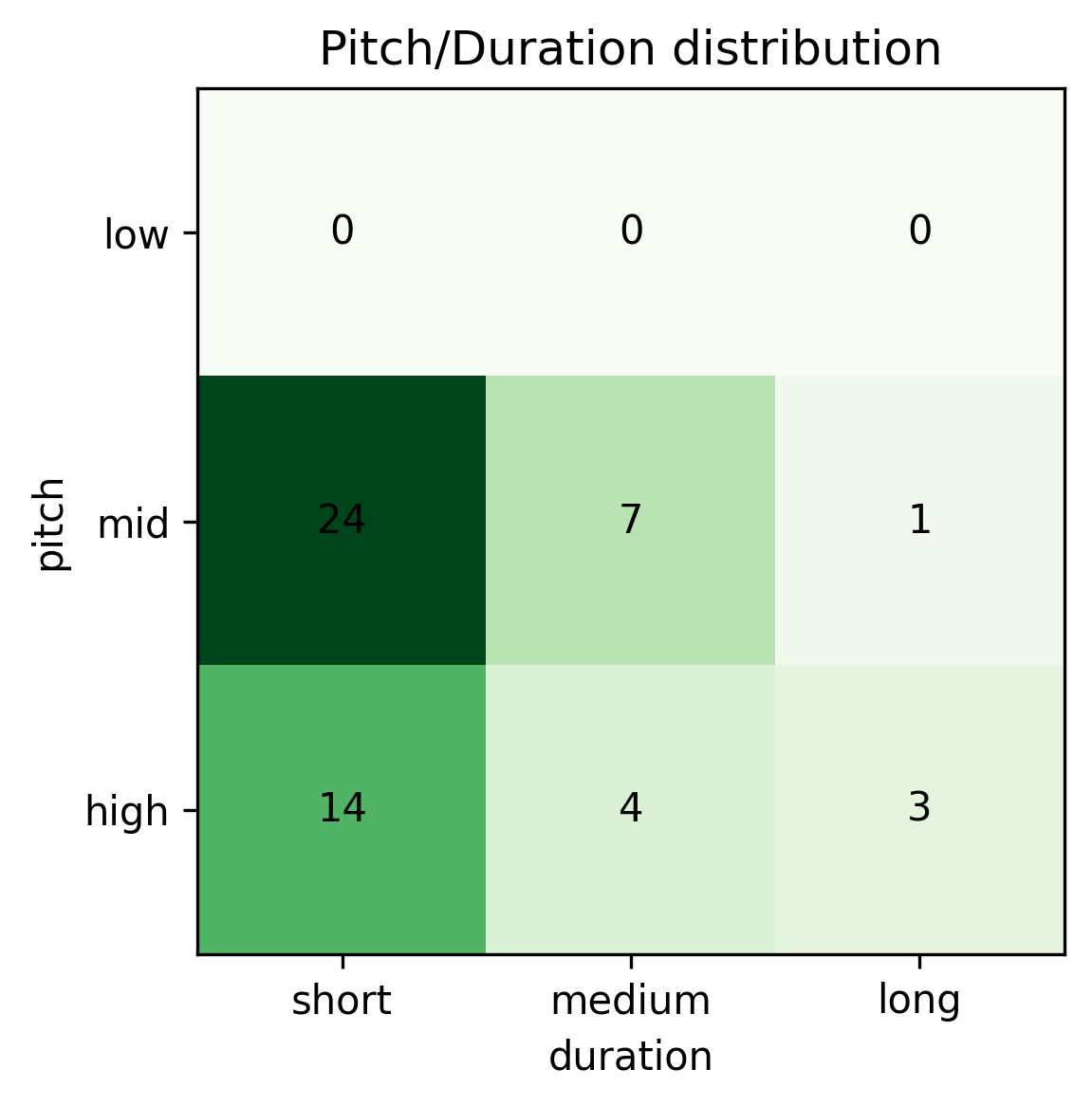}
            \caption{Variant B}
            \label{fig:var2}
        \end{subfigure}
        \hfill
        \begin{subfigure}[b]{0.32\textwidth}
            \centering
            \includegraphics[width=\textwidth]{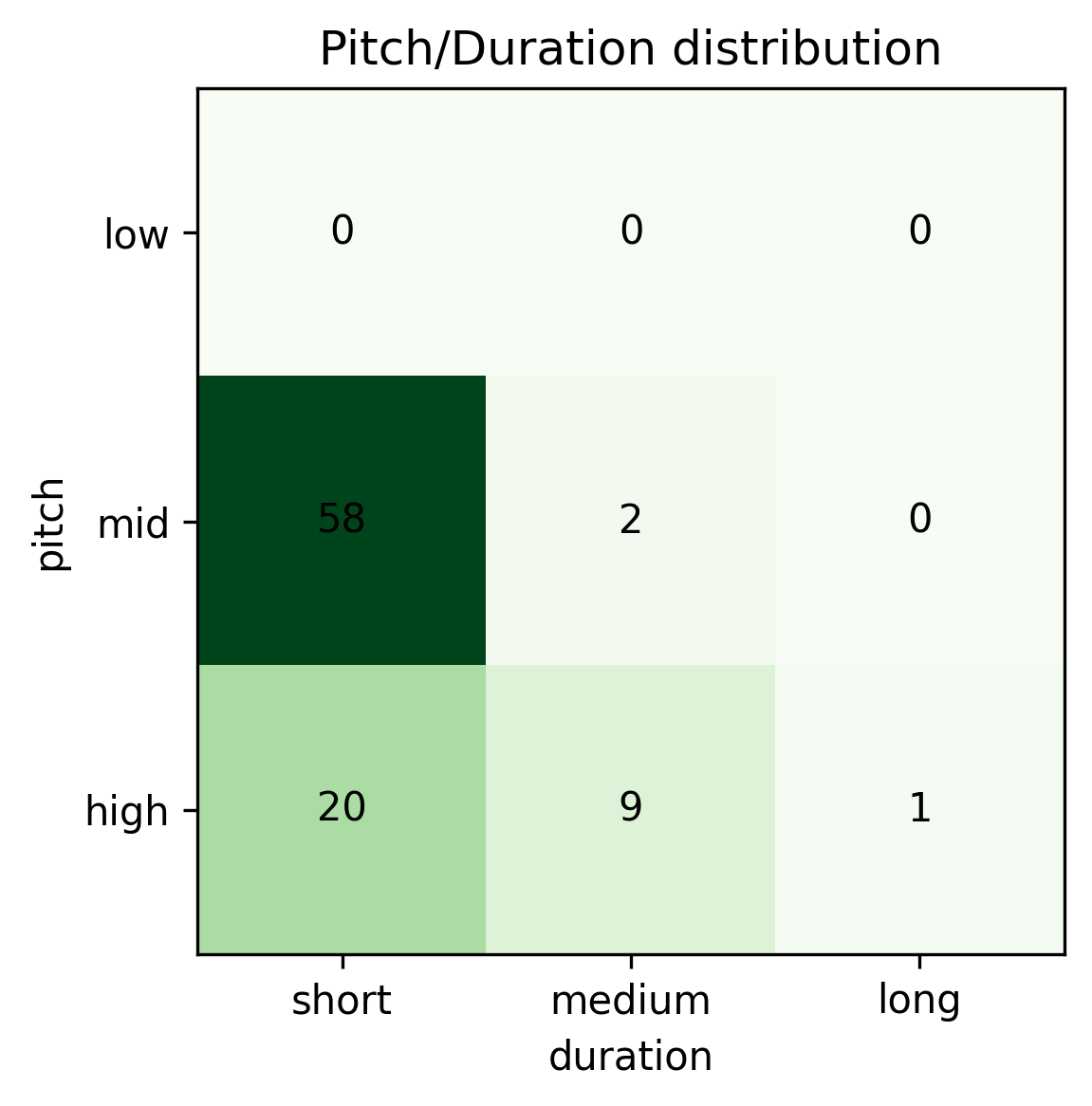}
            \caption{Variant C}
            \label{fig:var3}
        \end{subfigure}
        \caption{Individual pitch-duration distributions for the three randomly selected variants, highlighting differing compositional characters (lyrical, balanced, virtuosic).}
        \label{fig:Individual_PitchDurata}
    \end{figure}

    \textbf{Individual vs. aggregated frequency-duration distribution}: 
    Fig.~\ref{fig:Individual_PitchDurata} illustrates the profound rhythmic and melodic variability among the three randomly selected variants. The first variant (Fig.~\ref{fig:var1}) displays a balanced, potentially lyrical character, with occurrences evenly distributed between \texttt{short} (31) and \texttt{medium} (25) durations in the middle register. The second variant (Fig.~\ref{fig:var2}) shifts the focus toward higher pitches with rapid articulations (\texttt{high/short}, 14). Conversely, the third variant (Fig.~\ref{fig:var3}) exhibits a highly virtuosic or agitated pacing, characterized by a massive concentration of \texttt{short} durations in both the \texttt{mid} (58) and \texttt{high} (20) registers, with an almost complete absence of long notes.

\begin{figure}[h!]
        \centering
        \includegraphics[width=0.4\linewidth]{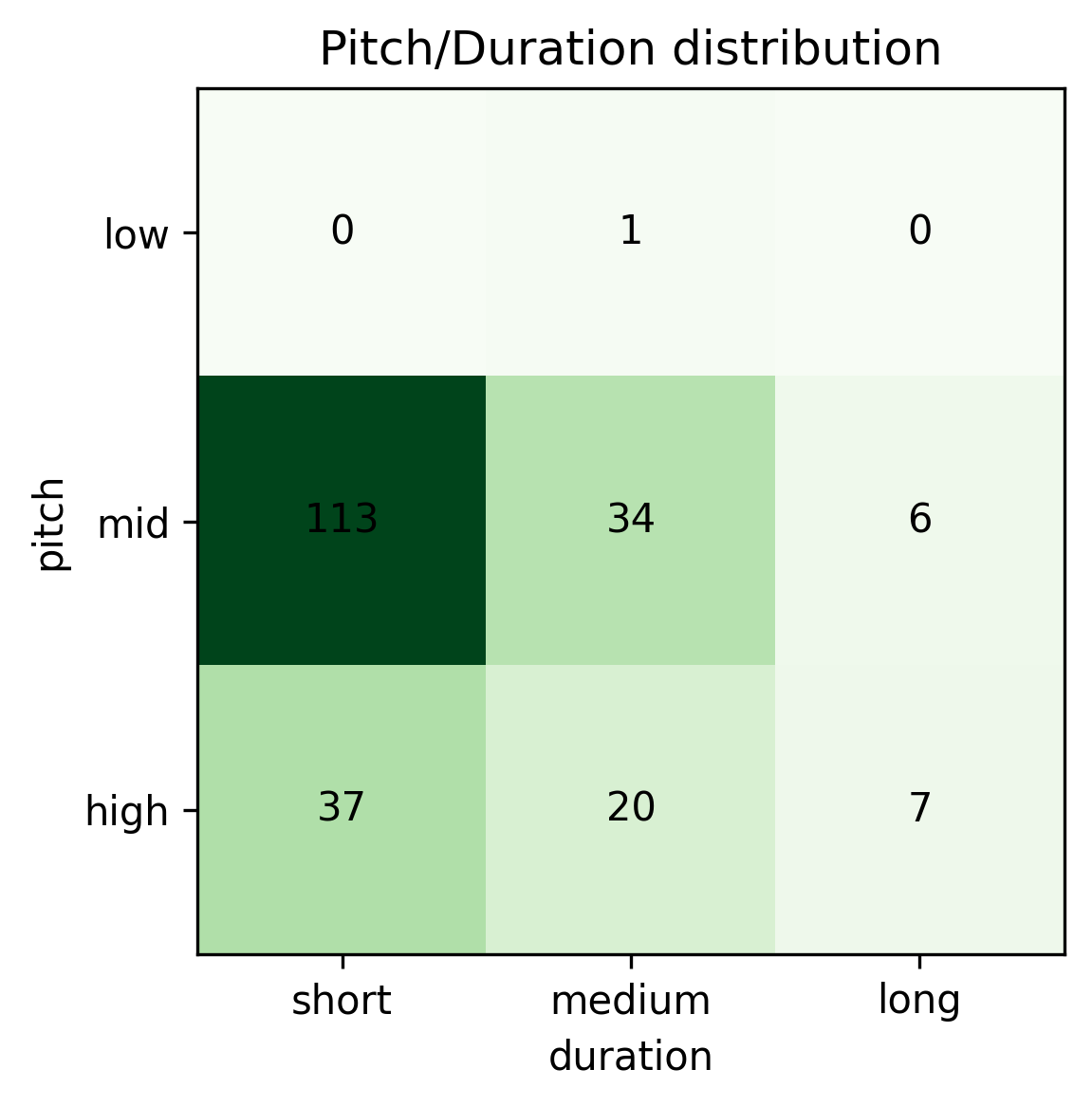} 
        \caption{Aggregated pitch-duration distribution for the selected subset of variants.}
        \label{fig:Aggregated_PitchDurata}
    \end{figure}

    Despite this extreme local variance, computing the aggregated distribution of the subset (Fig.~\ref{fig:Aggregated_PitchDurata}) smooths the outliers and reveals the foundational Rossinian macro-trend: the core melodic narrative unfolds predominantly in the middle register (\texttt{mid/short} dominates entirely with 113 occurrences, followed by \texttt{mid/medium} with 34). High-register peaks are treated systemically as structural exceptions.

The comparative analysis is applicable to every type of distribution of interest, as shown in the previous sections, and therefore extensible to any set of scores that can be the object of study.

\section{Conclusion and Future Works}

	The computational analysis of the corpus of authorial variants of Gioachino Rossini's ``Mi lagner\`o tacendo'', based on feature extraction via MusicXML parsing, structural representation through Graph Theory, and statistical distribution analysis, has allowed for the definition of a general methodology for the computational philological study of musical scores.
In this perspective, the Rossinian variants serve as a case study where musical elements are treated as historical-interpretative phenomena and as manifestations of a structured process. This process is formally described through similarity metrics and organizational configurations emerging directly from the data. This approach moves beyond purely qualitative analysis, offering systematic tools to identify recurrent compositional patterns, anomalous configurations within a single score, and degrees of structural proximity between different variants.
Furthermore, the formal structuring of the data facilitates the application of Machine Learning algorithms, such as Graph Neural Networks (GNNs), currently under investigation for both supervised classification of musical structures and unsupervised clustering for anomaly detection. 
In parallel, generative models for compositional prediction are in the simulation phase. These results confirm the potential for a coherent dialogue between scientific calculus, musicology, and machine learning, suggesting that a computational approach to large sets of scores can serve as a robust interpretative tool for investigating creative processes in musical practice.

\appendix
\section{Music Fundamentals}\label{Appendix - Music Fundamentals}

\begin{figure}[h!]
    \centering
    \begin{subfigure}[b]{0.4\textwidth}
        \centering
        \includegraphics[width=\linewidth]{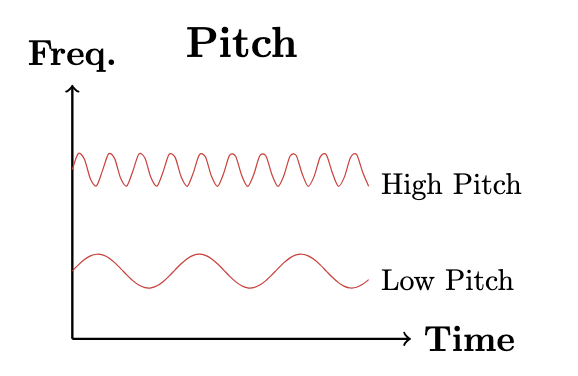}
        \caption{Pitch corresponds to frequency/height}
        \label{fig:pitch}
    \end{subfigure}
    \hfill 
    \begin{subfigure}[b]{0.4\textwidth}
        \centering
        \includegraphics[width=\linewidth]{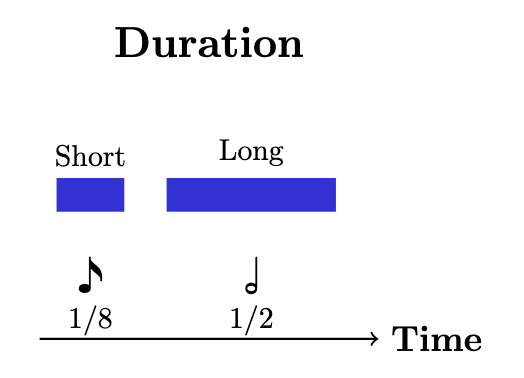}
        \caption{Duration corresponds to time length}
        \label{fig:duration}
    \end{subfigure}
    
    \caption{Music notation in a Cartesian System.}
    \label{fig:cartesian_system}
\end{figure}

To ensure analytical clarity, the fundamental musical variables employed in this study: \textit{Pitch}, \textit{Duration}, and \textit{Melodic Interval} are defined.
Following the conceptual framework described above, each musical score is modeled as a Cartesian system (Fig.~\ref{fig:cartesian_system}) mapping frequency against time with (``$H\!z$'' vs. ``$t$''). 

\begin{itemize}
    \item \textbf{Pitch (Vertical Axis/$H\!z$-coordinate\footnote{Mapped to standard MIDI (Musical Instrument Digital Interface) numbers (0-127 of classes of possible frequencies).}):} this represents the perceived ``height'' of a sound. Within our computational model, a higher position on the musical staff correlates with an increased frequency (acute sound), while a lower position corresponds to a lower frequency (grave sound).
    \item \textbf{Duration (Horizontal Axis/$t$-coordinate\footnote{Treated as quantized time units (e.g., quarters, eighths).}):} this represents how long a sound lasts in time.
\end{itemize}

To provide a complete reference for the difference between the concept of pitch (perceptual) and frequency (physical) reports the correspondences for the central octave \cite{Pitch}.
It is worth noting that while modern Western music typically uses Equal Tuning, historical contexts often refer to Pythagorean Tuning. Table \ref{Temperamenti} illustrates the theoretical divergence between these systems using the enharmonic pair $C\sharp4$ / $D\flat4$.

\begin{table}[h!]
\centering
    \caption{Correspondence between Pitch (note) and Frequency (Hz) in the central octave.}\label{pitchtofreq}
    \begin{tabular}{@{} l r @{}}
        \toprule
        \textbf{Note or Pitch} & \textbf{Frequency (Hz)} \\
        \midrule
        Do4 (C4)  & 261.63 \\
        $Re\flat4$ ($D\flat4$) & 277.18 \\
        Re4 (D4)  & 293.66 \\
        $Mi\flat4$ ($E\flat4$) & 311.13 \\
        Mi4 (E4)  & 329.63 \\
        Fa4 (F4)  & 349.23 \\
        $Sol\flat4$ ($G\flat4$) & 369.99 \\
        Sol4 (G4) & 392.00 \\
        $La\flat4$ ($A\flat4$) & 415.30 \\
        La4 (A4)  & 440.00 \\
        $Si\flat4$ ($B\flat4$) & 466.16 \\
        Si4 (B4)  & 493.88 \\
        \bottomrule
    \end{tabular}
\end{table}

\begin{table}[h!]
    \centering
    
    \begin{tabular}{||p{5cm}|p{3cm}|p{3.2cm}||}
        \hline
        \textbf{Note} & \textbf{Equal tuning (Hz)} & \textbf{Pythagorean tuning (Hz)} \\
        \hline
        $Do\sharp 4$ ($C\sharp{}4$) / $Re\flat4$ ($D\flat{}4$) & 277.18 & 281.00\footnotesize{*}  \\
        \hline
        \multicolumn{3}{l}
        
        {\footnotesize{* calculated as 277.18\,Hz\,$\times 2^{23.6/1200}\approx$\,281.00\,Hz}} \\
    \end{tabular}
    \caption{Intonation of $C\sharp4$ / $D\flat4$ in Equal vs. Pythagorean tuning.}
    \label{Temperamenti}
\end{table}

Furthermore, the movement from one note to the next one is defined as an Interval\footnote{Intervals can be Ascending or Descending.} (jump) $\Delta P$, 
mathematically defined as the absolute difference between the pitch values of two consecutive notes at time $t$ and $t+1$:
\begin{equation}
    \Delta P = |P_{t+1} - P_t|
\end{equation}
As an example of typology of analysis related to the intervals two types of leaps have been defined:
\begin{itemize}
    \item \textbf{Step:} $\Delta P \le 2$ semitones (distance $\le 1$ whole tone).
    \item \textbf{Leap:} $\Delta P > 2$ semitones (distance $> 1$ whole tone).
\end{itemize}

\newpage

\section*{Acknowledgements}
	The work of Drs. S. Licciardi and D. Macchione has been developed in the framework of the project ``EAR\_Enacting ARTISTIC RESEARCH -- WP2a'', code INTAFAM00060, CUP B83C24001590005, funded under the National Recovery and Resilience Plan (NRRP), Mission 4, by the European Union -- NextGenerationEU.\\
	The work of Prof. E. Francomano has been supported by ``MUR (Ministero dell'Universit\`a e della Ricerca) through the PNRR project ICON-Q, Partenariato Esteso NQSTI-PE00000023, Spoke 2`` and by Project GNCS 2025- INdAM. \\
	Thanks are due to the Kassel-based publishing house B\"arenreiter for making MLT's digital print proofs available for research.

\end{document}